\documentclass[12pt]{iopart}

\usepackage{cite}
\usepackage{xspace}
\usepackage{xcolor}
\usepackage{hyperref}
\usepackage{iopams}
\usepackage{graphicx}
\usepackage{booktabs}
\usepackage{dcolumn}

\begin{document}

\title[Impact of soft QCD effects over intrajet azimuthal anisotropies]{Impact of soft QCD effects over intrajet azimuthal anisotropies}

\author{Jesus Alberto V. Corral \footnote{Corresponding author} $^1$, A. Cota Rodriguez$^2$,
J. A. Murillo Quijada$^2$}
\address{$^1$ Departamento de F\'{i}sica, Universidad de Sonora}
\address{$^2$ Departamento de Investigaci\'{o}n en F\'{i}sica, Universidad de Sonora}

\ead{jesuscorralx@gmail.com, antonio.cota@unison.mx and javier.murillo@unison.mx}

\begin{abstract}

Observations of collective signatures within jets of particles have been recently reported by measurements at the Large Hadron Collider at CERN laboratory, providing insights on the minimum conditions required for forming a Quark Gluon Plasma state. A quantitative comparison with data measurements on intrajet azimuthal anisotropies and predictions from PYTHIA event generator soft QCD model parameter tunes: Monash, CP5, and QCD-scheme CR is presented. Energetic jets with transverse momentum over 550 GeV from proton-proton collisions at 13 TeV are reconstructed. Results from predictions using the standard E-scheme (ES) and jet axis choice are compared with those from the Winner-Take-All (WTA) frame. Significant differences are present between predictions from WTA and ES frames at the lowest and highest intrajet charged particles multiplicities (\Nch), with the QCD-scheme CR tune showing the smallest variation between frames. Predictions for \vTwo at the highest \Nch are consistent with the data measurements at a 2.3$\sigma$ uncertainty level. These results set new guidance towards the search for intrajet collective effects and understanding of nonperturbative QCD dynamics originated from a single parton.

\end{abstract}

\section{Introduction}
Minimal conditions that could lead to the strongly interacting state of matter known as Quark Gluon Plasma (QGP) are currently under debate. Its description in terms of a nearly ideal hydrodynamic behavior \cite{PhysRevD.46.229,Heinz:2013th,Gale:2013da} has been proposed as an alternative modelling strategy in terms of gluon saturation in the initial state~\cite{Dusling:2015gta, Nagle:2018eea}. QGP was first observed and studied in high energy nucleus-nucleus collisions (\ensuremath{A A}) ~\cite{add_rhic_AA:01,add_rhic_AA:04,cms:PbPbfirst,cms:PbPbsecond,ALICE:2011svq,ATLAS:2012at,star:AAfirst,star:AAsecond,phenix:AAfirst,STAR:oct2019,STAR:oct2018}. Collectivity signatures, usually present in the QGP state, were recently spotted to be present in smaller collision systems such as proton-proton (\PP) \cite{cms:ppfirst,Aad:2015gqa,cms:ppsecond,Khachatryan:2016txc,add_atlas_pp:01} and proton-nucleus \cite{cms:pPbfirst, add_rhic_pA:01,add_rhic_pA:02,Aad:2012gla,Aad:2013fja,Abelev:2012ola,Aaij:2015qcq,ABELEV:2013wsa,Khachatryan:2015waa,cms:pPbPbPb_corr_identifiedPar,Aaboud:2017acw,Aaboud:2017blb}. Probes for collective states and their interpretation suffered then a dramatic change of direction, for example by looking up for systems with even smaller interaction size. These systems, lately regarded as `small systems', are characterized by a smaller Number of Multiparton Interactions (nMPI)~\cite{nMPIs}. A larger magnitude of nMPI and produced particles are usually present in \ensuremath{A A} collisions, followed by \pPb and \PP collision systems, respectively. A recent study from the ALICE experiment shows evidence of collective effects and the appearance of the so-called `ridge' effect to very low events~\cite{ALICE:2023ulm}. The group of small systems under recent study includes the following collision species: electron-positron (\ensuremath{e^{+} e^{-}}\xspace), electron-proton, photon-proton, photon-nucleus (\ensuremath{\gamma A }\xspace)~\cite{alepCorr:2019, Belle:2022fvl, zeus:Dec2019_ep, ZEUS:2021qzg, CMS:2022doq, ATLAS:2021jhn} for which none of the measurements have revealed collective signatures apart from the \ensuremath{\gamma A }\xspace system.

Among the smallest systems that have been studied recently is the one that arises from an energetic single parton produced at high energy collisions. A single parton (either quark or gluon) that appears in final states in this environment experiences the so-called fragmentation process, which creates a cascade of partons known as `parton showering', followed by a strongly-coupled transition from partons into hadrons stage known as hadronization. The resulting collimated group of particles is known as a `jet' object. It has a cone shape as it arises from the single parton and slightly opens up angularly as particles travel away through the vacuum. Partons and their interactions are described by the theory of Quantum Chromodynamics (QCD). Phenomena involving low momentum transfer governed by the strong interaction, such as the QGP, the structure of hadrons, the formation of jets, or underlying activity in high energy collisions, are modelled with non-perturbative theoretical structures. These models own a set of free parameters that are adjusted so data modelling is as accurate as possible. Each combination of parameter values is regarded as a specific `tune', having unique patterns for observables from collision events. 

It has been recently postulated in~\cite{PhysRevC.107.064908} that nonperturbative QCD effects taking place during parton fragmentation could lead to collective effects within high multiplicity jets similar to those observed in \PP, \pPb and \PbPb systems. First studies with experimental data from the STAR and CMS experiments were reported in the following references~\cite{Putschke_2007, austin_CMS}. The latter reports a probe for collectivity signatures within high multiplicity jet objects that arise in \PP collisions at $\sqrt{s}$ $=$ 13 TeV. An enhancement of long-range elliptic anisotropies is observed with respect to the reference PYTHIA8 model with CP5 tune~\cite{CP5:2019} at the highest charged particle multiplicity inside the jet cone. This study aims to investigate the modelling accuracy of data measurements further, under the impact of soft QCD effects on the measurements of elliptic anisotropies as a function of charged particles multiplicity inside the jet cone, denoted as \Nch, especially in the region with larger multiplicities (\Nch $> 80$). An additional goal is the comparison of predictions from different jet reference frames and QCD model tunes from PYTHIA8 Monte Carlo event generator ~\cite{Bierlich:2022pfr}, which includes the phenomenological Lund string model for hadronization, introducing the origin of jets from multiparton interactions (MPI) and subsequent string fragmentation. This model has described the data at the Large Hadron Collider (LHC) successfully \cite{ATLAS:2024png}.

The paper is organized in the following way. Section~\ref{sec:tunes} outlines PYTHIA MC event generator soft QCD models. Section~\ref{sec:jet} explores the new frame definition and the choice of jet axis. Section~\ref{sec:level1} discusses specific signatures for search collectivity phenomena, azimuthal anisotropies. Section~\ref{sec:level2} explores the impact of tune choice over particle multiplicity. Section~\ref{sec:level3} presents the results using PYTHIA MC and comparison with recent data. The paper ends with a summary in Section~\ref{sec:level5}.

\normalsize

\section{\label{sec:tunes}PYTHIA soft QCD tunes}
For the PYTHIA8 soft QCD model, the tune choice is usually guided by data measurements from the LHC, Large Electron-Positron collider, Stanford Linear Collider, and (\ensuremath{e^{+} e^{-}}\xspace) experiments. This letter presents results using the PYTHIA simulation version 8.309 via three different tune choices introduced below. The original PYTHIA8 scheme, known as the tune Monash, incorporates a model for MPI and colour reconnection (CR), making it appropriate for general purpose event generations ~\cite{monash}. In contrast with the CMS precision tunes, the so-called CP5 tune is optimized for 13 TeV LHC data, which carries a significant improvement in the description of the underlying soft physics of pp collisions. Recently, this tune has been used for jet studies, such as in reference ~\cite{austin_CMS}, where its performance is compared with experimental data.

The CR mechanism is one of the most critical components of a tune since it rearranges the string connections of the Lund Model, shaping the hadronization stage, influencing particle multiplicities, and transverse momentum distributions, particularly in the context of small collision systems is crucial to modeling collective-like phenomena through nonperturbative QCD effects, as encapsulated in specific tunes ~\cite{Ortiz_Velasquez_2013}. In addition, the CR scheme has been worked out with a similar parametrized strategy by Monash and CP5 tunes, with the latter having an additional effect from fitting to  LHC 13 TeV data. A third tune known as `QCD-scheme CR' has been added to the comparisons, incorporating a set of colour rules following QCD theory. This tune, as employed in the Monash tune, retains the majority of parameters while introducing modifications specific to the CR implementation. This new model minimizes the string length according to a simplified version of the SU(3) color algebra. The model identifies all possible dipole pairs that can reconnect and selects configurations with shorter string lengths, and determines the probability that a reconnection is allowed~\cite{Christiansen_2015}.

PYTHIA8 Monash~\cite{monash}, CP5~\cite{CP5:2019}, and `QCD-scheme CR' tunes consist of specific values for relevant parameters such as the strong coupling $\alpha_S(M_Z)$, whose value influences the amount of QCD radiation generated by the PYTHIA algorithm ~\cite{Bierlich:2022pfr,Sjostrand}. These parameters affect the emission of Initial State Radiation (ISR), and Final State Radiation (FSR), specific parameters within the lund string model determining how the parton shower evolves or those affecting the probabilities of a certain flavor of mesons (light and heavy) and baryons being produced within a specific jet and color reconnections, the detailed discussion of such values can be found here ~\cite{monash,CP5:2019,Christiansen_2015} for each of the tunes respectively.

\section{\label{sec:jet}Jet frame and recombination scheme}
\begin{figure*}[h!]
\centering
\includegraphics[height=0.45\textwidth]{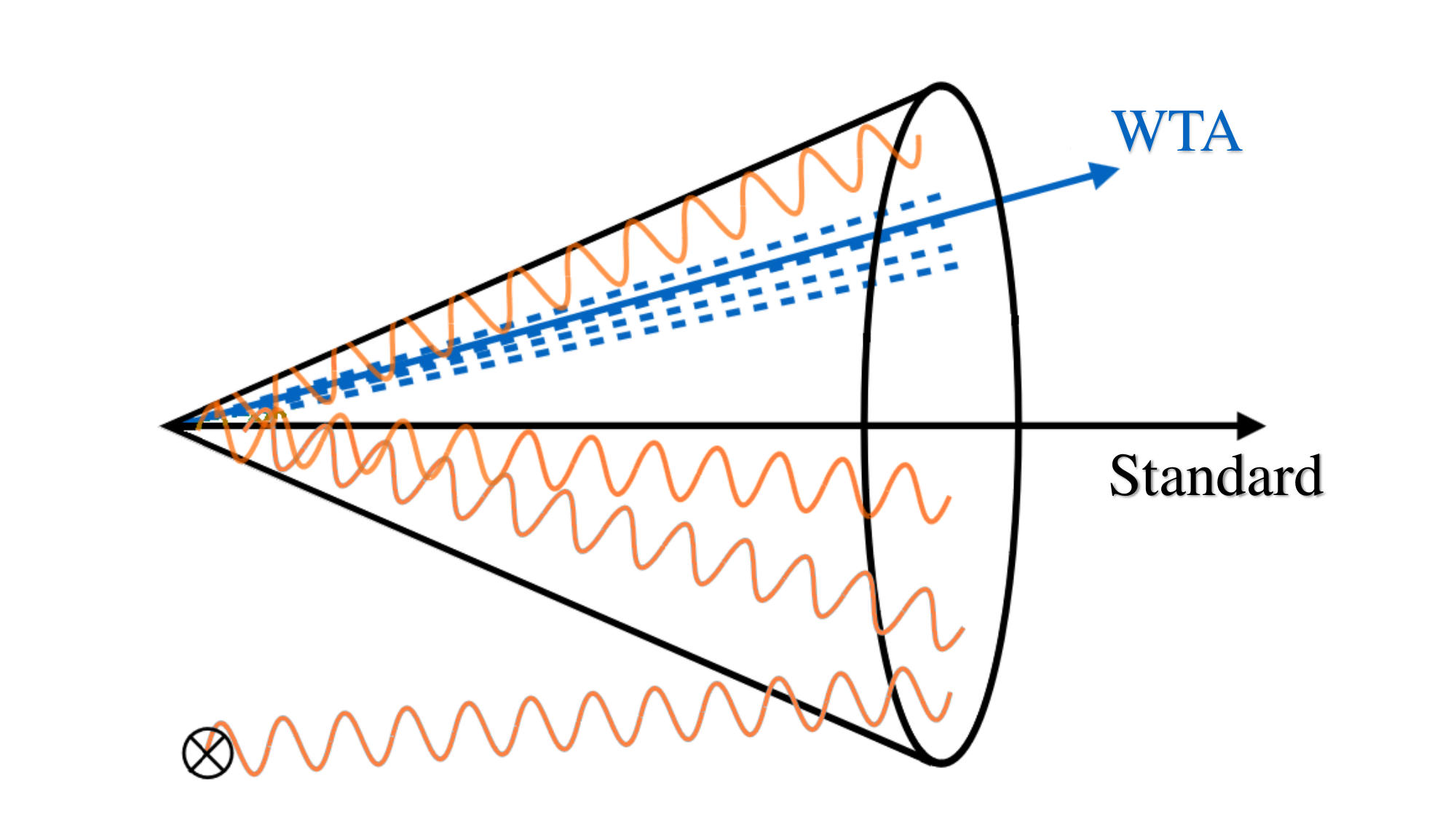}
\caption{\label{fig:schemes} Illustration to E-scheme jets axis (`standard') in black color along with WTA axis in blue color that is insensitive to soft radiation~\cite{Cal:2019gxa}.}
\end{figure*}

The analysis strategy follows the recent letter~\cite{austin_CMS}, which employs high transverse momentum (relative to the beam axis) jets reconstructed in an event using a particular algorithm with a choice of jet cone size. Aside from the QCD tune choice discussion, it is also relevant to investigate the effect of the choice of jet axis for the jet frame under consideration for intrajet measurements. The so-called `E-scheme' or `standard' recombination~\cite{Matteo} of jet reconstruction is used for jet studies with experimental data in~\cite{austin_CMS}, where the jet axis aligns with the summing of all particles within individual jets forming the jet momentum. For the `Winner-take-all' (WTA) scheme~\cite{Bertolini, Duff} a jet axis aligned to the harder particle is chosen, performing a pair-wise recombination that adds up the $p_{T}$ recursively but keeps the orientation of the particle with larger magnitude. This choice aims to minimize the effect of soft radiation recoils in the initial parton direction. Figure~\ref{fig:schemes} illustrates the jet collinear and soft radiation (blue dashed lines and orange curves, respectively), along with the orientation of standard (E-scheme) and WTA frames. Their separation is due to WTA following up collinear radiation, being insensitive to soft radiation.

For each case, a new coordinate frame system is established once the jet axis is defined. The momentum vectors of all particles within the jet cone are redefined for the new frame as $p^{*}=(j_{T},\eta^{*},\phi^{*})$. Here, $j_{T}$ denotes the transverse momentum of the particle for the selected jet axis. The origin of the $\phi^{*}$ is defined as $\hat{p}_{jet}\times $ $(\hat{p}_{jet} \times \hat{p}_{z})$, where $\hat{p}_{jet}$ is a vector along the jet momentum and $\hat{p}_{z}$ is the vector along the z axis. The $\eta^{*}$ is defined as $\eta^{*} = -ln[\tan{(\theta^{*}/2)}]$, where $\theta^{*}$ is taken with respect to the jet axis instead of the positive direction of the beam axis.

\section{\label{sec:level1}Azimuthal anisotropies}
The search for collectivity signatures can be done via characterizing two-particle azimuthal correlation distributions from particles that are far apart in rapidity. The 2-dimensional correlation distribution is usually built as in~\cite{CMS:2022doq} from angular differences $\DeltaEta$ and $\DeltaPhi$ corresponding to charged particle pairs in the collision event. The distribution is projected into the long range ($|\DeltaEta|>$2.0) and then fitted over the $\DeltaPhi$ range $[0, \pi]$ to a Fourier decomposition series $\propto 1 + \sum_n 2\VnDelta\cos(n\Delta\phi)$, from where the measured Fourier coefficients \VnDelta are extracted, where $n$ represents the order of the moment. The two-particle correlations are then factorized into the single-particle azimuthal anisotropy Fourier coefficients \vN as $\vN = \sqrt{\VnDelta}$ ~\cite{Voloshin:1996}. The second (\vTwo) and third (\vThree) coefficients are known as elliptic and triangular flow, respectively, and are directly related to the initial collision geometry and its fluctuations, which influence the medium evolution and provide information about its fundamental transport properties \cite{geometry_and_fluctuations_01,geometry_and_fluctuations_02,geometry_and_fluctuations_03,elliptical_triangular}. The \vN measurements with data can be contrasted with various models to extract conclusions on the presence of QGP or collective signatures.
\begin{figure*}[h!]
\centering
\includegraphics[height=0.65\textwidth]{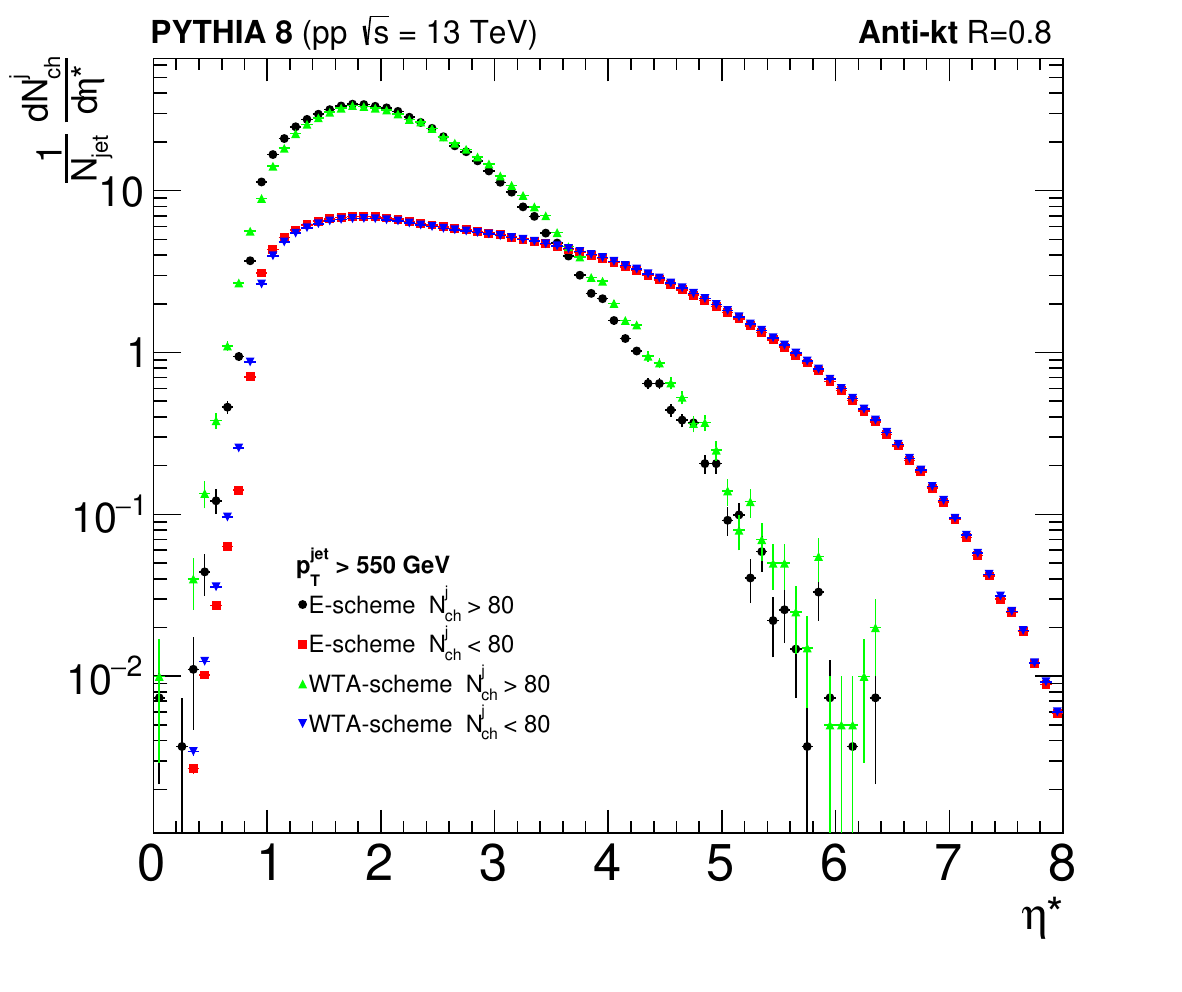}
\caption{\label{fig:etaprime_jetScheme}{The $\eta^*$ distributions of charged particles within the jet frame for E-scheme and WTA jets produced in \PP collisions at 13 TeV, The distributions consider jets with $p_{T}^{jet}$ $>$ 550 GeV and explores different jet multiplicity regimes, $\Nch$ $<$ 80 and  $\Nch$ $>$ 80. Normalized by number of jets ($N_{jet}$). These distributions are obtained using the Monash tune.}}
\end{figure*}

\section{\label{sec:level2}Data, simulation, and jet reconstruction}
To simulate jets originated from \PP collisions at $\sqrt{s}$ $=$ 13 TeV, the PYTHIA8 simulation version 8.309, and the soft QCD parameters for each of the three tunes described above were modified accordingly to generate samples with Monash, CP5, and QCD-scheme CR tunes. The FastJet package ~\cite{Cacciari:2011ma} for jet finding and reconstruction was used to implement the use of the anti-kt algorithm with jet cone size R= 0.8, selected jets are required to have $p_{T}^{jet}$ $>$ 550 GeV and pseudorapidity $|\eta^{jet}|$ $<$ 1.6 in the laboratory reference frame, their charged particles angular variables were transformed from laboratory values ($\eta$, $\phi$) to ($\eta^{*}, \phi^{*}$), to ensure consistency with experimental data, and taken their corresponding the two recombination scheme described above. Published data points from the following reference ~\cite{austin_CMS} that have been corrected by detector efficiency are included in plots to analyze the simulation modelling accuracy.

\begin{figure*}[h!]
    \centering
    \begin{minipage}[b]{0.49\textwidth}
        \centering
        \includegraphics[width=\textwidth]{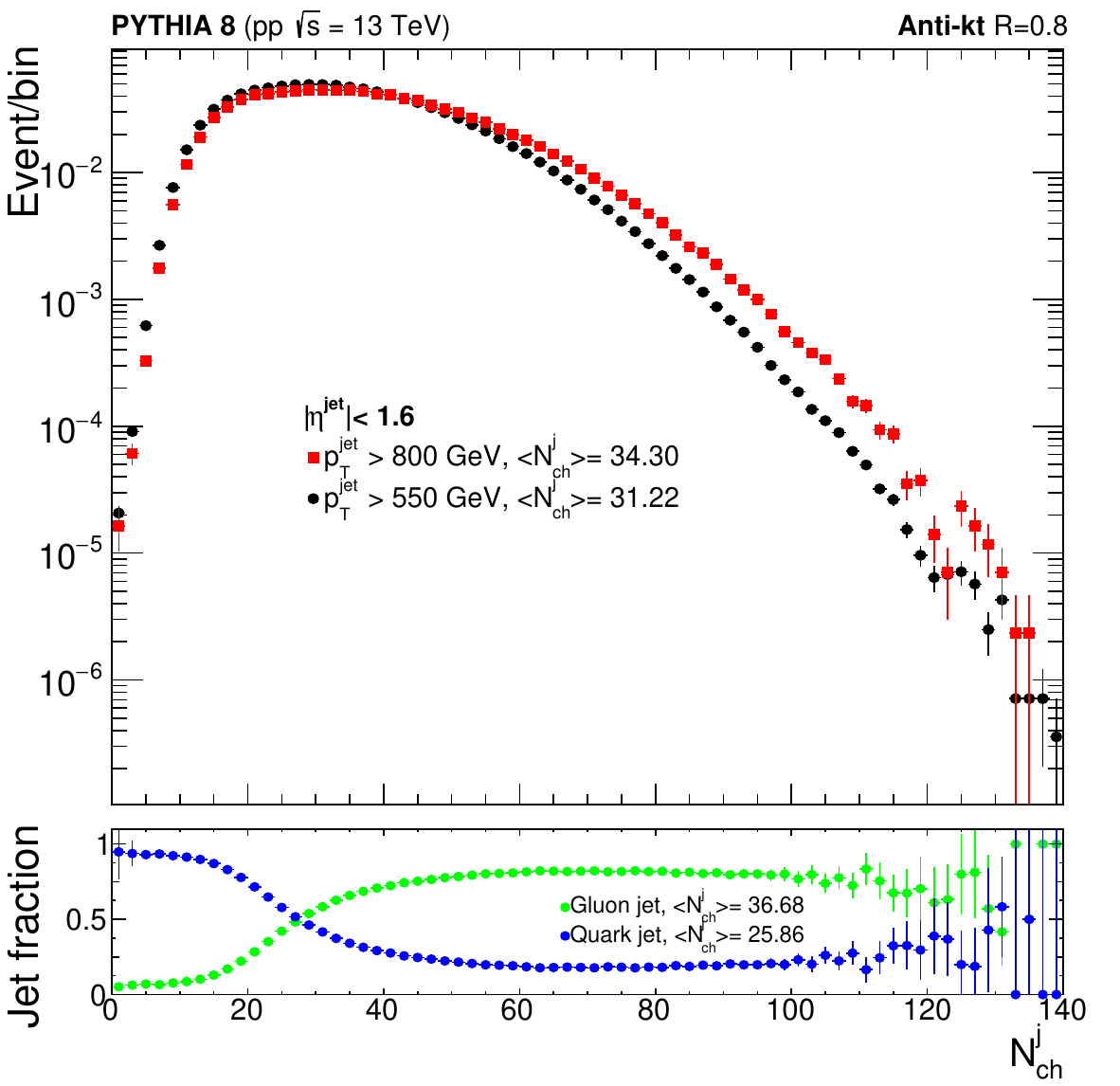}
        \caption{\label{fig:multiplicity1} $\Nch$ for jets produced in \PP colisions at 13 TeV with $p_{T}^{jet}>$ 550 GeV (black) and $p_{T}^{jet}>$ 800 GeV (red). Average \Nch values reflect the minimum energy threshold in each category. Normalized to unity.}
    \end{minipage}
    \hfill
    \begin{minipage}[b]{0.49\textwidth}
        \centering
        \includegraphics[width=\textwidth]{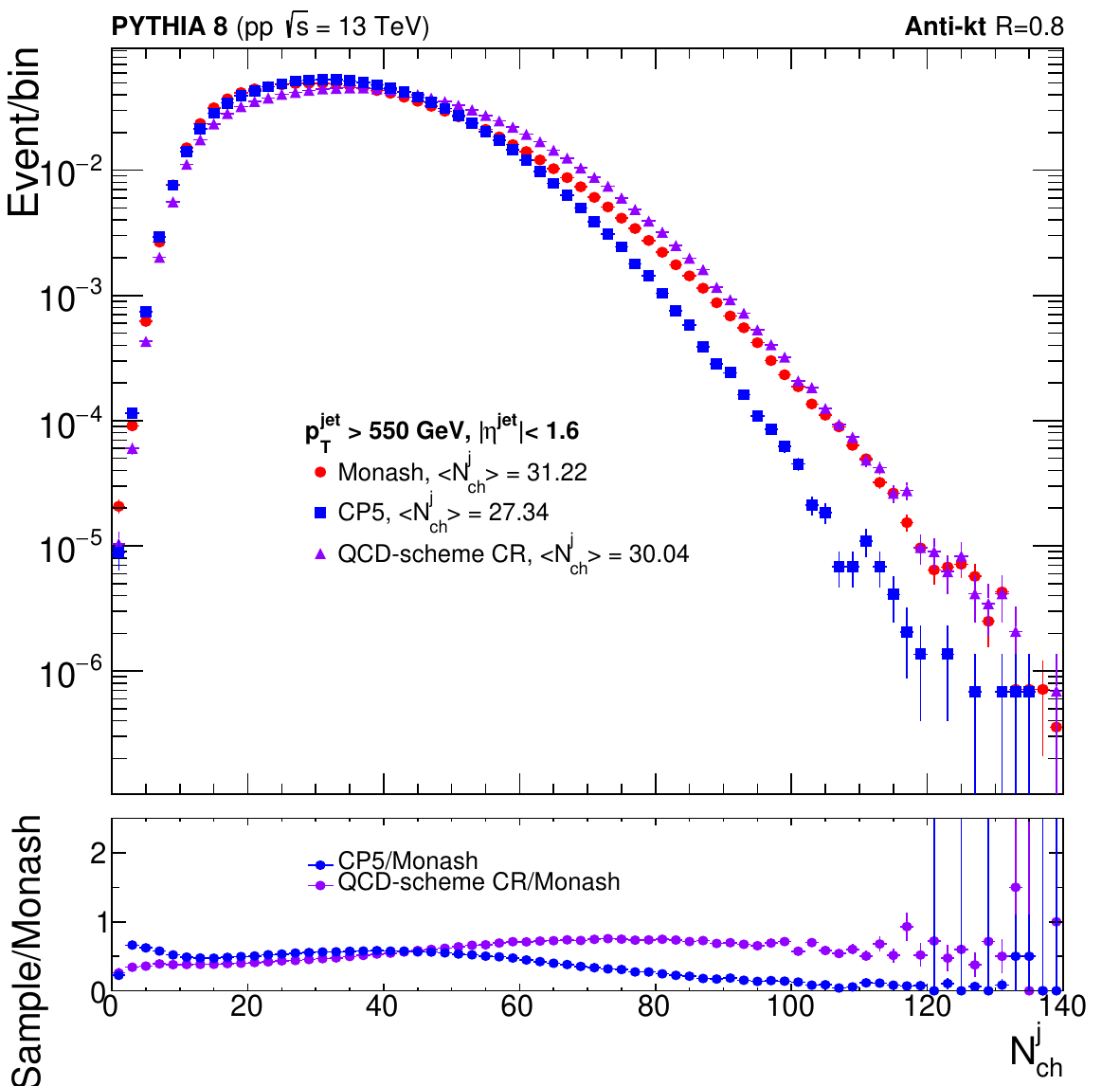}
        \caption{\label{fig:multiplicity_flavor} $\Nch$ for jets produced in \PP colisions at 13 TeV with $p_{T}^{jet}>$ 550 GeV. CP5 and QCD-scheme CR distributions are compared against Monash tune in the lower panel. Normalized to unity.}
    \end{minipage}
\end{figure*}

As in references~\cite{PhysRevC.107.064908, austin_CMS}, the reconstructed charged particle multiplicity (\Nch) is defined as the number of charged particles contained within the jet, with the requirement to have $p_{T} > 0.3$ GeV/$c$, and $|\eta|<2.4$ in the laboratory reference frame. Figure~\ref{fig:etaprime_jetScheme} includes $\eta^*$ distributions of charged particles within the jet in two different \Nch ranges below and above 80 and two jet reference systems. Events with lower charged particle multiplicity show a larger $\eta^*$ reach while the range is restricted up to $\sim$ 6 for \Nch$>$80. In this category, E-scheme shows a smaller $\eta^*$ average of 2.08 with respect 2.12 in WTA-scheme. Figure~\ref{fig:multiplicity1} shows the \Nch spectra from selected jets using the Monash tune. The mean value of charged particle multiplicity inside the jet cone, denoted as \Ntrackavg, is consistent with what is shown in~\cite{PhysRevC.107.064908}, with a range restricted up to \Nch $\sim$140. The \Ntrackavg value corresponding to selected jets with transverse momentum $p_{T}^{jet}>$ 550 GeV and $p_{T}^{jet}>$ 800 GeV is 31.22 and 34.30, respectively. 

\begin{table*}[h!]
\centering
\caption{Mean \Nch for different multiplicity classes. Statistical uncertainties are negligible.}
\begin{tabular}{lccc}
\hline
 & Monash & CP5 & QCD-scheme CR \\ 
\hline
$2  \leq$ \Nch $< 20$ & 15.45 & 14.64 & 14.81  \\
$20  \leq$ \Nch $< 32$ & 26.45 & 25.32 & 25.44  \\
$32  \leq$ \Nch $< 44$ & 37.94 & 36.62 & 36.91 \\
$44  \leq$ \Nch $< 56$ & 49.52 & 48.15 & 48.51  \\ 
$56  \leq$ \Nch $< 68$ & 61.22 & 59.86 & 60.25 \\ 
$68  \leq$ \Nch $< 80$ & 73.00 & 71.19 & 71.47  \\ 
$80  \leq$ \Nch $< 90$ & 84.35 & 81.09 & 81.61  \\ 
$90  \leq$ \Nch $< 100$ & 94.29 & 90.83 & 92.01  \\ 
 \Nch $\geq 100 $& 106.5 & 102.5 & 106.2  \\ 
Inclusive & 31.22 & 27.34 & 30.04  \\ 
\hline
\end{tabular}
\label{tab:MeanMult} 
\end{table*}

The increase in jet energy is then correlated with the capability for particle production. Furthermore, the distributions can be separated by jet origin parton species, either a gluon or a quark. The panel below in the same figures shows the fraction of these two origin categories for the $p_{T}^{jet}>$ 550 GeV sample, as this selection is used for the data measurements shown in the following sections. This comparison shows that for \Nch $\geq$25, jets originated by gluons dominate in fraction and quark originated jets dominate for the lowest multiplicities (\Nch $< 25$). Figure~\ref{fig:multiplicity_flavor} shows a comparison for \Nch prediction from Monash, CP5, and QCD-scheme CR tunes for the jet sample with $p_{T}^{jet}>$ 550 GeV. Shapes are comparable with each other within the  2 $<$ \Nch $\gtrsim$ 40 region. This consistency is kept so far between Monash and QCD-scheme CR tunes for the highest multiplicities, but displays a clear discrepancy with respect to CP5 prediction, which shows a lower yield. Table~\ref{tab:MeanMult} shows \Ntrackavg numbers for different \Nch categories for the three tunes. The inclusive average is indicated in the last row. Rows for $90  \leq$ \Nch $< 100$  and \Nch $\geq  100$ categories, confirm lower \Ntrackavg values for the CP5 category concerning Monash and QCD-scheme CR tunes.

\section{\label{sec:level3}Results}
\subsection{Long range Fourier decomposition comparison to data}
The two-particle correlation analysis techniques described below are identical to those described in Section~\ref{sec:level1}, but consider only charged particles within the reconstructed jet object, as previously discussed in reference ~\cite{austin_CMS}.  For each jet with $p_{T}^{jet}$ $> 550$ GeV, the two-dimensional (2D) angular correlation function is calculated using charged particles with $0.3 < j_{T} < 3.0$ \GeVc. The measurements are separated in \Nch classes. For each multiplicity class (i.e., a specific bin in \Nch), the trigger particles are particles whose $j_{T}$, labeled as $j_{T}^{trig}$, is within a particular range, in this case $0.3 < j_{T} < 3.0$  \GeVc. These are correlated angularly with other associated particles $j_{T}^{assos}$, which are also selected within the same $j_{T}$ range $0.3 < j_{T} < 3.0$ \GeVc. This selection ensures full consistency with the experimental methodology outlined in reference~\cite{austin_CMS}.

The $\DeltaEta^*$ and $\DeltaPhi^*$ are the differences in $\eta^*$ and $\phi^*$ of each charged particle pair taken using the E-scheme frame. The two-dimensional (2D) correlation function, defined as in reference~\cite{PhysRevC.107.064908}, is projected onto the $\DeltaPhi^*$ axis for $|\DeltaEta^*| > 2$, excluding short-range correlations. Consistent with experimental reference, the particles with $\eta^{*}>5$ are excluded in the subsequent analysis.

\begin{figure*}[h!]
\centering
\includegraphics[height=0.33\textwidth]{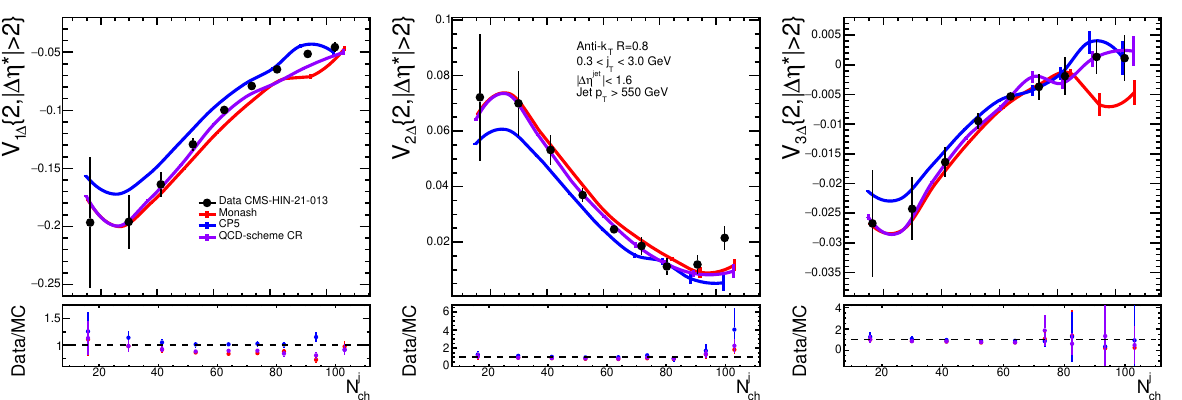}
\caption{\label{fig:VnDelta_nch}Fitted \VnDelta coefficients as a function of \Nch. \VoneDelta, \VtwoDelta, and \VthreeDelta coefficient distributions are shown in the left, central, and right plots, respectively. Vertical axis ranges have been adjusted on the lower panel so all points are visible in each case.}
\end{figure*}

A Fourier decomposition series fit as described in Section~\ref{sec:level1} is performed to extract intrajet $\VnDelta$ coefficients. Figure~\ref{fig:VnDelta_nch} shows the dependence of  \VoneDelta, \VtwoDelta, and \VthreeDelta coefficients as a function of \Nch for the three QCD tunes along with data measurements from reference~\cite{austin_CMS}. Only the first three terms are included in the fit since additional terms have a negligible effect on its quality. Each point corresponds to a specific \Nch category with ascending values. The point with the highest multiplicity corresponds to \Nch $\geq 100$ selection. Lower panels display the comparison of different tunes against data measurements, for \VoneDelta distribution, the uncertainties in both data and simulation are comparable, making the ratio uncertainties visible in the full \Nch range, where Monash and QCD-scheme CR tunes show in general a better agreement with data for the \Nch $< 80$ region, with slightly better modelling from the latter, with a clear difference spotted for data from CP5 tune that for most points go beyond the uncertainty limit. For \VtwoDelta distributions, the situation is similar to that with \VoneDelta, with a visible data enhancement over all tune predictions in the highest \Nch point, for this region, Monash and QCD-scheme CR tunes values show a significant consistency with data measurements, better than CP5. The \VthreeDelta distribution, QCD-scheme CR generally maintains the most consistent agreement with data for all multiplicities. However, the uncertainty for \VtwoDelta and \VthreeDelta at low \Nch are comparable with data, but at high \Nch the uncertainties increase, making them appear visually negligible for low \Nch range.

\subsection{Single-particle anisotropy and jet axis WTA frame}

\begin{figure*}[h!]
\includegraphics[height=0.55\textwidth]{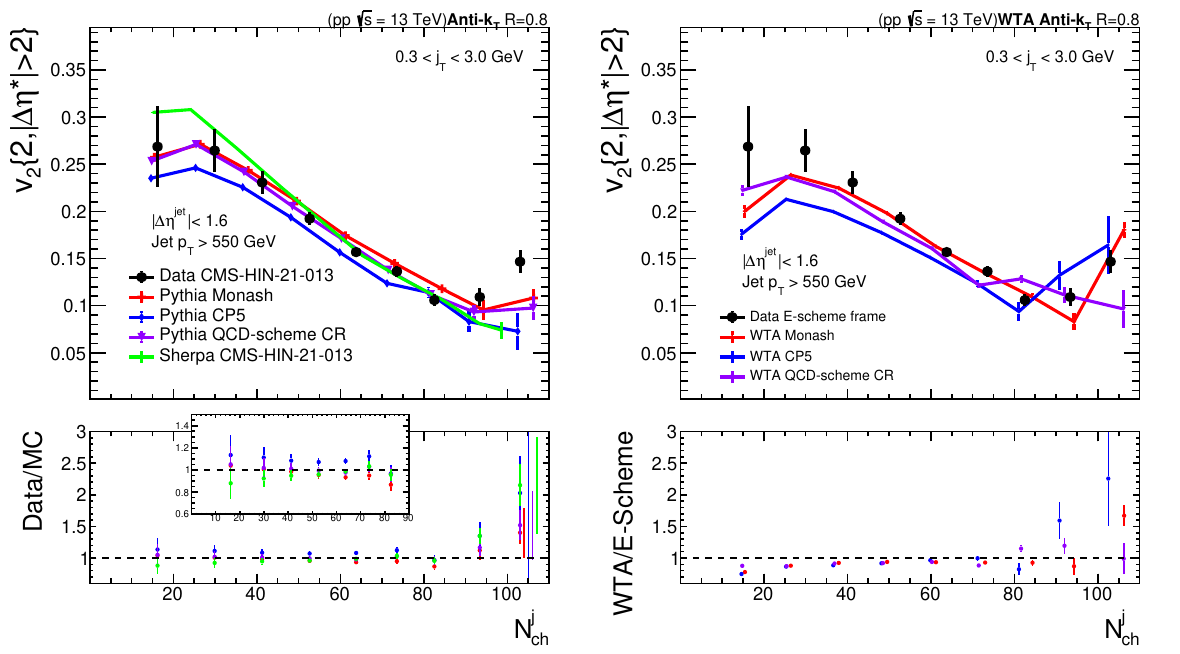}
\caption{\label{fig:v2Delta_nch} \vTwo anisotropy coefficient as a function of \Nch using E-scheme (left) and WTA scheme (right). Data measurement points from reference~\cite{austin_CMS} are only available for the E-scheme. Ratios are shown at the lower panel, and the left lower panel for this last bin with 1$\sigma$ and 2.3$\sigma$ uncertainty limits. A special pad has been added to display better the comparison in the central region with 10$ <$\Nch $< 90$ where better consistency is spotted for QCD-scheme CR tune.}
\end{figure*}

At the left of Figure~\ref{fig:v2Delta_nch}, the single-particle anisotropy \vTwo as a function of \Nch is shown. Prediction from the three tunes is included in Figure~\ref{fig:v2Delta_nch} along with the data measurements from~\cite{austin_CMS}. An additional reference from the Sherpa event generator is added from the same reference~\cite{austin_CMS}. QCD-scheme CR and Monash tunes are visibly in better agreement with the data measurements, with slightly better modelling from QCD-scheme CR in the \Nch $< 80$ region, in general, present better modelling than CP5 tune from \vTwo values with respect to data for all \Nch range. This is quantified in Table~\ref{tab:chi2_results} in the first column, $\chi^2$ comparison test results are shown for \Nch $< 80$, where QCD-scheme CR tune displays the lower value of 0.32 below the Sherpa reference and Monash with 1.02 and 2.28, respectively. CP5 tune displays a larger distance to data with respect to Monash and QCD-scheme CR. This behaviour is consistent with what is shown in the second column in Table~\ref{tab:chi2_results}, listing a lower $\chi^2$ value for the Monash tune that is a factor of 2 lower than the values for CP5 and Sherpa references shown in ~\cite{austin_CMS} measurements. For the largest multiplicity category, as with \VtwoDelta in Figure~\ref{fig:VnDelta_nch}, in Figure ~\ref{fig:v2Delta_nch}, for \vTwo anisotropy the data measurements show an excess over the prediction, Monash and QCD-scheme CR tunes are yet consistent with the data distribution at 2.3 sigma confidence level. These limits are extracted from experimental \vTwo value and its uncertainty from CMS reference and MC tune predictions, ratios are shown in the lower left panel of the same Figure for this last bin with 1$\sigma$ and 2.3$\sigma$ uncertainty limits.

\begin{table*}
    \centering
    \caption{Chi-squared values for $\chi^2/$ndof comparison of different models.}
    \begin{tabular}{l c c}
        \hline
        \textbf{Model} & \textbf{$\chi^2/ndof$ ($\Nch$ $< 80$)} & \textbf{$\chi^2/ndof$ 
        ($\Nch$ $\geq 80$)} \\
        \hline
        Monash & 11.41 / 5 = 2.28 & 20.40 / 2 = 10.20 \\
        CP5    & 23.44 / 5 = 4.69 & 47.10 / 2 = 23.55 \\
        QCD-scheme CR  & 1.61 / 5 = 0.32 & 20.41 / 2 = 10.21 \\
        CMS-HIN-21-013 Sherpa & 5.11 / 5 = 1.02 & 43.85 / 2 = 21.93 \\
        \hline
    \end{tabular}
    \label{tab:chi2_results}
\end{table*}

At the right in the same Figure the effect over \vTwo elliptical coefficient from transforming intrajet charged particles $\DeltaEta^*$, $\DeltaPhi^*$ coordinates into the WTA frame is shown as a function of \Nch. Here, the lower panel shows a comparison of each tune prediction with respect to their values from the E-scheme frame on the left. The \vTwo values from both frames are so far consistent with each other for most of the range within \Nch $< 80$ category, with smaller values for WTA, especially at the lowest \Nch limits. QCD-scheme CR shows stronger overall consistency for results in both frames, especially at the largest \Nch $\geq 80$ region. Monash results are comparable with those from QCD-scheme CR except for the region with the highest multiplicity, where Monash gets a significantly larger value with respect to its E-scheme prediction. CP5 shows the largest discrepancy from its corresponding E-scheme values in this region. The observed discrepancy between schemes at the largest \Nch for Monash tune is a consequence of a smaller $\eta^*$ average in E-scheme with respect to the WTA shown in Figure~\ref{fig:etaprime_jetScheme}, which affects the $\DeltaEta^*$ values that are projected to the $\DeltaPhi^*$ distribution.

\section{\label{sec:level5}Summary}
Recent probes for collective effects within energetic jets with transverse momentum \pTjet $>$ 550 GeV produced from proton-proton collisions have been performed at the LHC. In these measurements, enhancement of long-range elliptic \vTwo anisotropies is observed with respect to the reference PYTHIA8 model with CP5 tune~\cite{cp5} at the highest \Nch. This probe has motivated an inspection of the performance of different soft QCD tunes from PYTHIA simulation, namely: Monash as the original PYTHIA8 scheme, QCD-scheme CR tune as a contrast to the CP5 tune predictions used in experimental probes. Monash and QCD-scheme CR tunes display a better modelling of the data measurements for \VnDelta and \vTwo coefficients in the \Nch $< 80$ region.

These results indicate that QCD-scheme CR tune modelling adds more theoretical considerations within SU(3) QCD color algebra, apart from other soft QCD effects are essential for an expanded data modelling, in particular on \vTwo anisotropy measurements throughout full \Nch range. The parametrized CR model contained within Monash is already enough to give reasonable data modelling with less accuracy. Both are followed by a CP5 scheme that is further tuned from jet studies to high energy LHC data. Differences in performance between Monash and CP5 tunes could be attributed to variations in the colour reconnection range parameter, which impacts the probability of reconnecting parton systems. CP5 was specially tuned to improve ISR, FSR, and MPI effects so the difference could also be attributed to other soft QCD effects. Additionally, differences could originate from the strong coupling associated with multiple parton interactions.

Enhancement of data for the \vTwo coefficient in \Nch $\geq 80$ region over tunes predictions has been spotted, though QCD-scheme CR tune is consistent with the data measurements at 2.3$\sigma$ confidence level; at the very edge for claiming such excess as significant and introduces motivation for further measurements with larger available datasets from the LHC.

The effect of varying the jet frame to WTA frame has turned into consistent predictions from all tunes for \Nch $< 80$ region concerning their corresponding E-scheme values in most of the range, except for very low \Nch values, where WTA delivers smaller anisotropy magnitudes for all tunes. QCD-scheme CR tune shows a better overall consistency, followed by Monash and CP5, respectively, in \Nch $\geq 80$ region, with the latter having larger values with respect to its E-scheme prediction, in general, all tunes exhibit discrepancies between the WTA and E-scheme frames, particularly at the lowest \Nch and highest ends of the available \Nch range. Since the WTA frame minimizes the effect of soft radiation recoils in the initial parton direction, it is of great importance to add up the data measurements for WTA frame.

These results contribute to the understanding of the collective effect origin and non-perturbative dynamics of multiparton systems at the smaller scales by inspecting with larger detail the effect of the tune setup and jet frame used in the measurements. Particularly, the results allow us to conclude that the PYTHIA8 QCD tune choice has a significant effect over intrajet azimuthal measurements. Specifically, QCD-scheme CR and Monash tunes are found to be more suitable choices that display better modelling to data measurements. Given this, any difference between those tunes will be attributed to the different CR mechanisms. As predictions from these two tunes at the highest \Nch are located at the border of the 2$\sigma$ uncertainty level concerning the data measurements, a more detailed experimental probe with finer \Nch resolution is necessary to declare precise conclusions on the existence of intrajet collectivity. Similarly, considering the three tunes under test deliver different \vTwo predictions at the highest \Nch after rotating into the WTA frame, it is crucial to have an experimental reference to these predictions to extend the scope of the conclusions concerning the effect of soft radiation over charged particles pseudorapidities.

\section*{Data availability statement}
All data that support the findings of this study are included within the article (and any supplementary files).

\section*{References}
\bibliography{apssamp}

\providecommand{\noopsort}[1]{}\providecommand{\singleletter}[1]{#1}%
\providecommand{\newblock}{}
\begin{thebibliography}{10}
\expandafter\ifx\csname url\endcsname\relax
  \def\url#1{{\tt #1}}\fi
\expandafter\ifx\csname urlprefix\endcsname\relax\def\urlprefix{URL }\fi
\providecommand{\eprint}[2][]{\url{#2}}

\bibitem{PhysRevD.46.229}
Ollitrault J~Y 1992 {\em Phys. Rev. D\/} {\bf 46}(1) 229

\bibitem{Heinz:2013th}
Heinz U and Snellings R 2013 {\em Ann. Rev. Nucl. Part. Sci.\/} {\bf 63} 123
  (\textit{Preprint} \eprint{1301.2826})

\bibitem{Gale:2013da}
Gale C, Jeon S and Schenke B 2013 {\em Int. J. Mod. Phys. A\/} {\bf 28} 1340011
  (\textit{Preprint} \eprint{1301.5893})

\bibitem{Dusling:2015gta}
Dusling K, Li W and Schenke B 2016 {\em Int. J. Mod. Phys. E\/} {\bf 25}
  1630002 (\textit{Preprint} \eprint{1509.07939})

\bibitem{Nagle:2018eea}
Nagle J~L and Orjuela~Koop J 2019 {\em Nucl. Phys. A\/} {\bf 982} 455
  (\textit{Preprint} \eprint{1807.04619})

\bibitem{add_rhic_AA:01}
Ackermann K~H {\em et~al.\/} (STAR) 2001 {\em Phys. Rev. Lett.\/} {\bf 86} 402
  (\textit{Preprint} \eprint{nucl-ex/0009011})

\bibitem{add_rhic_AA:04}
Alver B {\em et~al.\/} (PHOBOS) 2010 {\em Phys. Rev. C\/} {\bf 81} 024904
  (\textit{Preprint} \eprint{0812.1172})

\bibitem{cms:PbPbfirst}
Chatrchyan S {\em et~al.\/} (CMS) 2011 {\em JHEP\/} {\bf 10} 076
  (\textit{Preprint} \eprint{1105.2438})

\bibitem{cms:PbPbsecond}
Chatrchyan S {\em et~al.\/} (CMS) 2012 {\em Eur. Phys. J. C\/} {\bf 72} 10052
  (\textit{Preprint} \eprint{1201.3158})

\bibitem{ALICE:2011svq}
Aamodt K {\em et~al.\/} (ALICE) 2012 {\em Phys. Lett. B\/} {\bf 708} 249--264
  (\textit{Preprint} \eprint{1109.2501})

\bibitem{ATLAS:2012at}
Aad G {\em et~al.\/} (ATLAS) 2012 {\em Phys. Rev. C\/} {\bf 86} 014907
  (\textit{Preprint} \eprint{1203.3087})

\bibitem{star:AAfirst}
Abelev B~I {\em et~al.\/} (STAR) 2009 {\em Phys. Rev. C\/} {\bf 80} 064912
  (\textit{Preprint} \eprint{0909.0191})

\bibitem{star:AAsecond}
Abelev B~I {\em et~al.\/} (STAR) 2010 {\em Phys. Rev. Lett.\/} {\bf 105} 022301
  (\textit{Preprint} \eprint{0912.3977})

\bibitem{phenix:AAfirst}
Adcox K {\em et~al.\/} (PHENIX) 2005 {\em Nucl. Phys. A\/} {\bf 757} 184
  (\textit{Preprint} \eprint{nucl-ex/0410003})

\bibitem{STAR:oct2019}
Adam J {\em et~al.\/} (STAR) 2020 {\em Phys. Rev. C\/} {\bf 101} 014916
  (\textit{Preprint} \eprint{1906.09204})

\bibitem{STAR:oct2018}
Adam J {\em et~al.\/} (STAR) 2018 {\em Phys. Lett. B\/} {\bf 783} 459
  (\textit{Preprint} \eprint{1803.03876})

\bibitem{cms:ppfirst}
Khachatryan V {\em et~al.\/} (CMS) 2010 {\em JHEP\/} {\bf 09} 091
  (\textit{Preprint} \eprint{1009.4122})

\bibitem{Aad:2015gqa}
Aad G {\em et~al.\/} (ATLAS) 2016 {\em Phys. Rev. Lett.\/} {\bf 116} 172301
  (\textit{Preprint} \eprint{1509.04776})

\bibitem{cms:ppsecond}
Khachatryan V {\em et~al.\/} (CMS) 2015 {\em Phys. Rev. Lett.\/} {\bf 116}
  172302 (\textit{Preprint} \eprint{1510.03068})

\bibitem{Khachatryan:2016txc}
Khachatryan V {\em et~al.\/} (CMS) 2017 {\em Phys. Lett. B\/} {\bf 765} 193
  (\textit{Preprint} \eprint{1606.06198})

\bibitem{add_atlas_pp:01}
Aad G {\em et~al.\/} (ATLAS) 2020 {\em Phys. Rev. Lett.\/} {\bf 124} 082301
  (\textit{Preprint} \eprint{1909.01650})

\bibitem{cms:pPbfirst}
Chatrchyan S {\em et~al.\/} (CMS) 2013 {\em Phys. Lett. B\/} {\bf 718} 795
  (\textit{Preprint} \eprint{1210.5482})

\bibitem{add_rhic_pA:01}
Aidala C {\em et~al.\/} (PHENIX) 2017 {\em Phys. Rev. C\/} {\bf 95} 034910
  (\textit{Preprint} \eprint{1609.02894})

\bibitem{add_rhic_pA:02}
Aidala C {\em et~al.\/} (PHENIX) 2019 {\em Nature Phys.\/} {\bf 15} 214
  (\textit{Preprint} \eprint{1805.02973})

\bibitem{Aad:2012gla}
Aad G {\em et~al.\/} (ATLAS) 2013 {\em Phys. Rev. Lett.\/} {\bf 110} 182302
  (\textit{Preprint} \eprint{1212.5198})

\bibitem{Aad:2013fja}
Aad G {\em et~al.\/} (ATLAS) 2013 {\em Phys. Lett. B\/} {\bf 725} 60
  (\textit{Preprint} \eprint{1303.2084})

\bibitem{Abelev:2012ola}
Abelev B {\em et~al.\/} (ALICE) 2013 {\em Phys. Lett. B\/} {\bf 719} 29
  (\textit{Preprint} \eprint{1212.2001})

\bibitem{Aaij:2015qcq}
Aaij R {\em et~al.\/} (LHCb) 2016 {\em Phys. Lett. B\/} {\bf 762} 473
  (\textit{Preprint} \eprint{1512.00439})

\bibitem{ABELEV:2013wsa}
Abelev B~B {\em et~al.\/} (ALICE) 2013 {\em Phys. Lett. B\/} {\bf 726} 164
  (\textit{Preprint} \eprint{1307.3237})

\bibitem{Khachatryan:2015waa}
Khachatryan V {\em et~al.\/} (CMS) 2015 {\em Phys. Rev. Lett.\/} {\bf 115}
  012301 (\textit{Preprint} \eprint{1502.05382})

\bibitem{cms:pPbPbPb_corr_identifiedPar}
Khachatryan V {\em et~al.\/} (CMS) 2015 {\em Phys. Lett. B\/} {\bf 742} 200
  (\textit{Preprint} \eprint{1409.3392})

\bibitem{Aaboud:2017acw}
Aaboud M {\em et~al.\/} (ATLAS) 2017 {\em Eur. Phys. J. C.\/} {\bf 77} 428
  (\textit{Preprint} \eprint{1705.04176})

\bibitem{Aaboud:2017blb}
Aaboud M {\em et~al.\/} (ATLAS) 2018 {\em Phys. Rev. C\/} {\bf 97} 024904
  (\textit{Preprint} \eprint{1708.03559})

\bibitem{nMPIs}
Collaboration 2022 {\em P\/} \urlprefix\url{h}

\bibitem{ALICE:2023ulm}
Acharya S {\em et~al.\/} (ALICE) 2024 {\em Phys. Rev. Lett.\/} {\bf 132} 172302
  (\textit{Preprint} \eprint{2311.14357})

\bibitem{alepCorr:2019}
Badea A, Baty A, Chang P, Innocenti G~M, Maggi M, Mcginn C, Peters M, Sheng
  T~A, Thaler J and Lee Y~J 2019 {\em Phys. Rev. Lett.\/} {\bf 123} 212002
  (\textit{Preprint} \eprint{1906.00489})

\bibitem{Belle:2022fvl}
Chen Y~C {\em et~al.\/} (Belle) 2022 {\em Phys. Rev. Lett.\/} {\bf 128} 142005
  (\textit{Preprint} \eprint{2201.01694})

\bibitem{zeus:Dec2019_ep}
Abt I {\em et~al.\/} (ZEUS) 2020 {\em JHEP\/} {\bf 04} 070 (\textit{Preprint}
  \eprint{1912.07431})

\bibitem{ZEUS:2021qzg}
Abt I {\em et~al.\/} (ZEUS) 2021 {\em JHEP\/} {\bf 12} 102 (\textit{Preprint}
  \eprint{2106.12377})

\bibitem{CMS:2022doq}
Tumasyan A {\em et~al.\/} (CMS) 2023 {\em Phys. Lett. B\/} {\bf 844} 137905
  (\textit{Preprint} \eprint{2204.13486})

\bibitem{ATLAS:2021jhn}
Aad G {\em et~al.\/} (ATLAS) 2021 {\em Phys. Rev. C\/} {\bf 104} 014903
  (\textit{Preprint} \eprint{2101.10771})

\bibitem{PhysRevC.107.064908}
Baty A, Gardner P and Li W 2023 {\em Phys. Rev. C\/} {\bf 107}(6) 064908
  \urlprefix\url{https://link.aps.org/doi/10.1103/PhysRevC.107.064908}

\bibitem{Putschke_2007}
Putschke J and (forthe STAR~Collaboration) 2007 {\em Journal of Physics G:
  Nuclear and Particle Physics\/} {\bf 34} S679
  \urlprefix\url{https://dx.doi.org/10.1088/0954-3899/34/8/S72}

\bibitem{austin_CMS}
Collaboration C 2023 {\em Physics Review Letters (PRL)\/}
  \urlprefix\url{https://doi.org/10.48550/arXiv.2312.17103}

\bibitem{CP5:2019}
Sirunyan A~M {\em et~al.\/} (CMS) 2020 {\em Eur. Phys. J. C\/} {\bf 80} 4
  (\textit{Preprint} \eprint{1903.12179})

\bibitem{Bierlich:2022pfr}
Bierlich C {\em et~al.\/} 2022 {\em SciPost Phys. Codeb.\/} {\bf 2022} 8
  (\textit{Preprint} \eprint{2203.11601})

\bibitem{ATLAS:2024png}
Aad G {\em et~al.\/} (ATLAS) 2024  (\textit{Preprint} \eprint{2405.20206})

\bibitem{monash}
Skands P, Carrazza S and Rojo J 2014 {\em Eur. Phys. J. C\/} {\bf 74} 3024

\bibitem{Ortiz_Velasquez_2013}
Ortiz~Velasquez A, Christiansen P, Cuautle~Flores E, Maldonado~Cervantes I~A
  and Pai\ifmmode~\acute{c}\else \'{c}\fi{} G 2013 {\em Phys. Rev. Lett.\/}
  {\bf 111}(4) 042001
  \urlprefix\url{https://link.aps.org/doi/10.1103/PhysRevLett.111.042001}

\bibitem{Christiansen_2015}
Christiansen J~R and Skands P~Z 2015 {\em Journal of High Energy Physics\/}
  {\bf 2015} ISSN 1029-8479
  \urlprefix\url{http://dx.doi.org/10.1007/JHEP08(2015)003}

\bibitem{Sjostrand}
Sjostrand T and Skands P~Z 2005 {\em Eur. Phys. J. C\/} {\bf 39} 129--154
  (\textit{Preprint} \eprint{hep-ph/0408302})

\bibitem{Matteo}
Cacciari M, Salam G~P and Soyez G 2008 {\em Journal of High Energy Physics\/}
  {\bf 2008} 063
  \urlprefix\url{https://dx.doi.org/10.1088/1126-6708/2008/04/063}

\bibitem{Bertolini}
Bertolini~Daniele Chan~Tucker T~J 2014 {\em Journal of High Energy Physics\/}
  {\bf 2014} 1029--8479 \urlprefix\url{https://doi.org/10.1007/JHEP04(2014)013}

\bibitem{Duff}
Neill~Duff T~J 2014 {\em Journal of High Energy Physics\/} {\bf 2014}
  1029--8479 \urlprefix\url{https://doi.org/10.1007/JHEP04(2014)017}

\bibitem{Cal:2019gxa}
Cal P, Neill D, Ringer F and Waalewijn W~J 2020 {\em JHEP\/} {\bf 04} 211
  (\textit{Preprint} \eprint{1911.06840})

\bibitem{Voloshin:1996}
Voloshin S and Zhang Y 1996 {\em Phys. Rev. C.\/} {\bf 70} 665
  (\textit{Preprint} \eprint{hep-ph/9407282})

\bibitem{geometry_and_fluctuations_01}
Alver B~H, Gombeaud C, Luzum M and Ollitrault J~Y 2010 {\em Phys. Rev. C\/}
  {\bf 82} 034913 (\textit{Preprint} \eprint{1007.5469})

\bibitem{geometry_and_fluctuations_02}
B~Schenke S~Jeon C~G 2011 {\em Phys. Rev. Lett\/} {\bf 106} 042301
  (\textit{Preprint} \eprint{1009.3244})

\bibitem{geometry_and_fluctuations_03}
Z~Qiu C~S and Heinz U 2012 {\em Phys. Lett. B\/} {\bf 707} 151
  (\textit{Preprint} \eprint{1110.3033})

\bibitem{elliptical_triangular}
Alver B and Roland G 2010 {\em Phys. Rev. C\/} {\bf 81} 054905 [Erratum: Phys.
  Rev. C 82 (2010) 039903] (\textit{Preprint} \eprint{1003.0194})

\bibitem{Cacciari:2011ma}
Cacciari M, Salam G~P and Soyez G 2012 {\em Eur. Phys. J. C\/} {\bf 72} 1896
  (\textit{Preprint} \eprint{1111.6097})

\bibitem{cp5}
Collaboration C 2023 {\em The European Physical Journal C\/}
  \urlprefix\url{https://doi.org/10.1140/epjc/s10052-023-11630-8}

\end{thebibliography}

\end{document}